\documentclass{appolb}
\usepackage{epsfig}
\def\lsim{\raise0.3ex\hbox{$<$\kern-0.75em\raise-1.1ex\hbox{$\sim$}}}
\def\gsim{\raise0.3ex\hbox{$>$\kern-0.75em\raise-1.1ex\hbox{$\sim$}}}

\begin{document}
\title{\bf QCD thermodynamics in the crossover/freeze-out region
}
\author{Frithjof Karsch
\address{Fakult\"at f\"ur Physik, Universit\"at Bielefeld, D-33615 Bielefeld,
Germany\\
and\\
Physics Department, Brookhaven National Laboratory, Upton, NY 11973, USA}
}
\maketitle
\begin{abstract}
We use results from a $6^{th}$ order Taylor expansion of the QCD equation
of state to construct expansions for cumulants
of conserved charge fluctuations and their correlations. 
We show that these cumulants strongly constrain the range of
applicability of hadron resonance gas model calculations. We point 
out that the latter is inappropriate to describe equilibrium properties 
of QCD at zero and non-zero values of the baryon
chemical potential already at $T\sim 155$ MeV.
\end{abstract}
  
\section{Introduction}
The existence of a critical point in the phase diagram of strong interaction
matter will be clearly visible in the (singular) structure of 
higher order cumulants of net baryon number, strangeness and electric charge
fluctuations and their cross correlations.
A prerequisite for utilizing these 
theoretically well founded observables in experimental searches for 
the possible existence of a critical point is that experimentally 
observed charge fluctuations
are indeed generated close to the pseudo-critical line that 
characterizes the chiral crossover transition and eventually ends in 
the critical point. 

Establishing the relation between freeze-out conditions at different
beam energies in heavy ion experiments and the crossover line in QCD
thus is of paramount importance. Although strategies have been developed
to extract freeze-out conditions from observables that are directly
accessible to experiment as well as calculations within QCD
\cite{Karsch:2010ck,Karsch:2012wm,Bazavov:2012vg}, applying these in practice
is still hampered by large statistical errors and our poor control over 
systematic effects entering the
measurement of higher order cumulants of charge fluctuations. The still
preferred approach to extract freeze-out conditions from experimental
data thus proceeds through a comparison of experimental data for particle
yields with model calculations based on the thermodynamics of a hadron 
resonance gas (HRG) \cite{Adamczyk:2017iwn,Floris:2014pta}. 

 
We will confront here some QCD calculations of higher order cumulants
of net charge fluctuations with corresponding HRG model calculations
and discuss the range of applicability of the latter.

\section{QCD thermodynamics at non-zero net baryon-number density}
\begin{figure}[t]
\begin{center}
\hspace*{-0.4cm}\includegraphics[width=72mm]{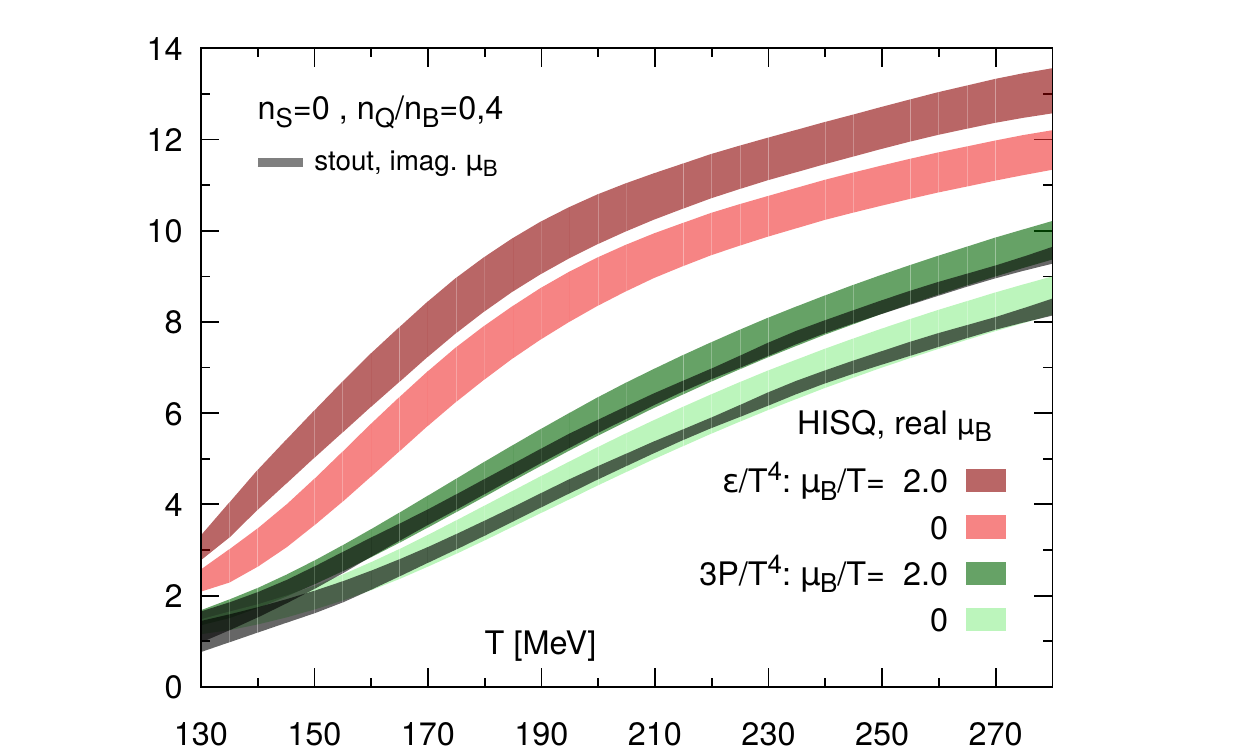}\hspace*{-1.2cm}
\includegraphics[width=70mm]{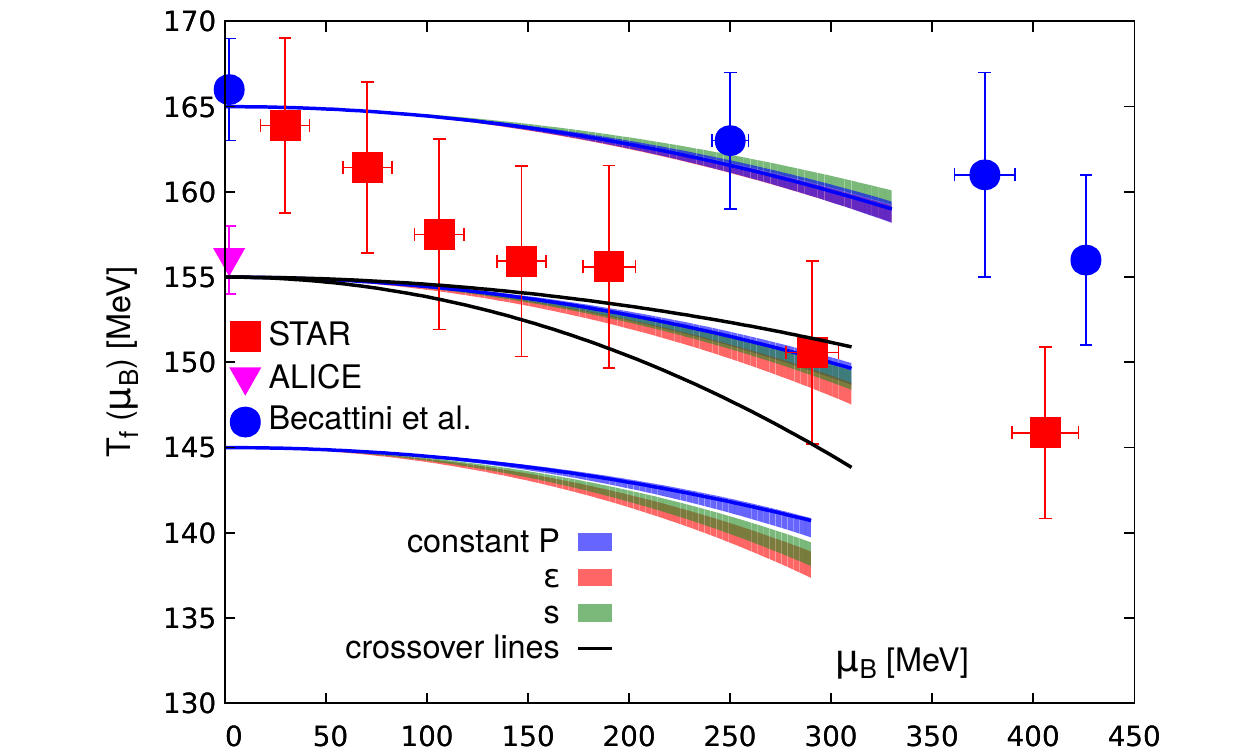}
\caption{{\it Left:}
The pressure and energy density versus temperature for two values of the
baryon chemical potential for strangeness neutral systems ($n_S=0$)
and a fixed ratio of net electric charge and net baryon-number
($n_Q/n_B=0.4$). Shown are results from a $6^{th}$ order Taylor expansion
\cite{Bazavov:2017dus} and from simulations with an imaginary chemical
potential \cite{Gunther:2016vcp}.
\newline
{\it Right:}
Lines of constant pressure,
energy density and entropy density versus baryon chemical potential
in (2+1)-flavor QCD for three different initial sets of values fixed at
$\mu_B=0$ and $T_0=145$~MeV, $155$~MeV and $165$~MeV, respectively.
}
\label{fig:LCP}
\end{center}
\end{figure}

The equation of state (EoS) of QCD with physical light and strange quark
masses has been analyzed at non-zero values of the baryon number
($\mu_B$), strangeness ($\mu_S$) and electric charge ($\mu_Q$) chemical 
potentials. For the case
of systems with vanishing net strangeness, $n_S=0$, and 
a fixed ratio on net electric charge to baryon number densities, $n_Q/n_B=0$, 
the EoS has been calculated in a $6^{th}$ order Taylor expansion
\cite{Bazavov:2017dus} as well as in simulations with an imaginary chemical 
potential \cite{Gunther:2016vcp}. These results
agree well among each other and are considered to be reliable up to  
$\mu_B/T\simeq 2$. Results from these 
calculations are shown in Fig.~\ref{fig:LCP}~(left).

Using the Taylor series for pressure ($P$), energy $(\epsilon)$ and 
entropy ($s$) density,
lines of constant physics (LCPs) can be determined,
\begin{equation}
T_f(\mu_B) = T_0 \left(1-\kappa_2^f 
\left( \frac{\mu_B}{T_0}\right)^2- \kappa_4^f \left( \frac{\mu_B}{T_0}
\right)^4\right) +{\cal O}(\mu_B^6) \; ,
\label{Tf}
\end{equation}
with $f$ labeling the observable that is kept constant, $f=P,\ \epsilon$ or $s$.
It turns out that up to $\mu_B\simeq 2T$
the correction arising from the quartic term is small. In the crossover
region, $145~{\rm MeV}\le T\le 165~{\rm MeV}$, the quadratic expansion 
coefficients, $\kappa_2^f$, 
vary in a range $0.006\le \kappa_2^f\le 0.012$ \cite{Bazavov:2017dus}. 
These LCPs can be compared with results for the pseudo-critical temperature 
of the chiral transition, $T_c(\mu_B)$, which can be parametrized as in
Eq.~\ref{Tf}. The spread of curvature coefficients determined for the chiral
crossover line\footnote{For a list of recent references see, for instance,
\cite{Bazavov:2017dus}}, $T_c(\mu_B)$, agrees well with
those obtained for the LCPs. 

\begin{figure}[t]
\begin{center}
\hspace*{-0.3cm}\includegraphics[width=68mm]{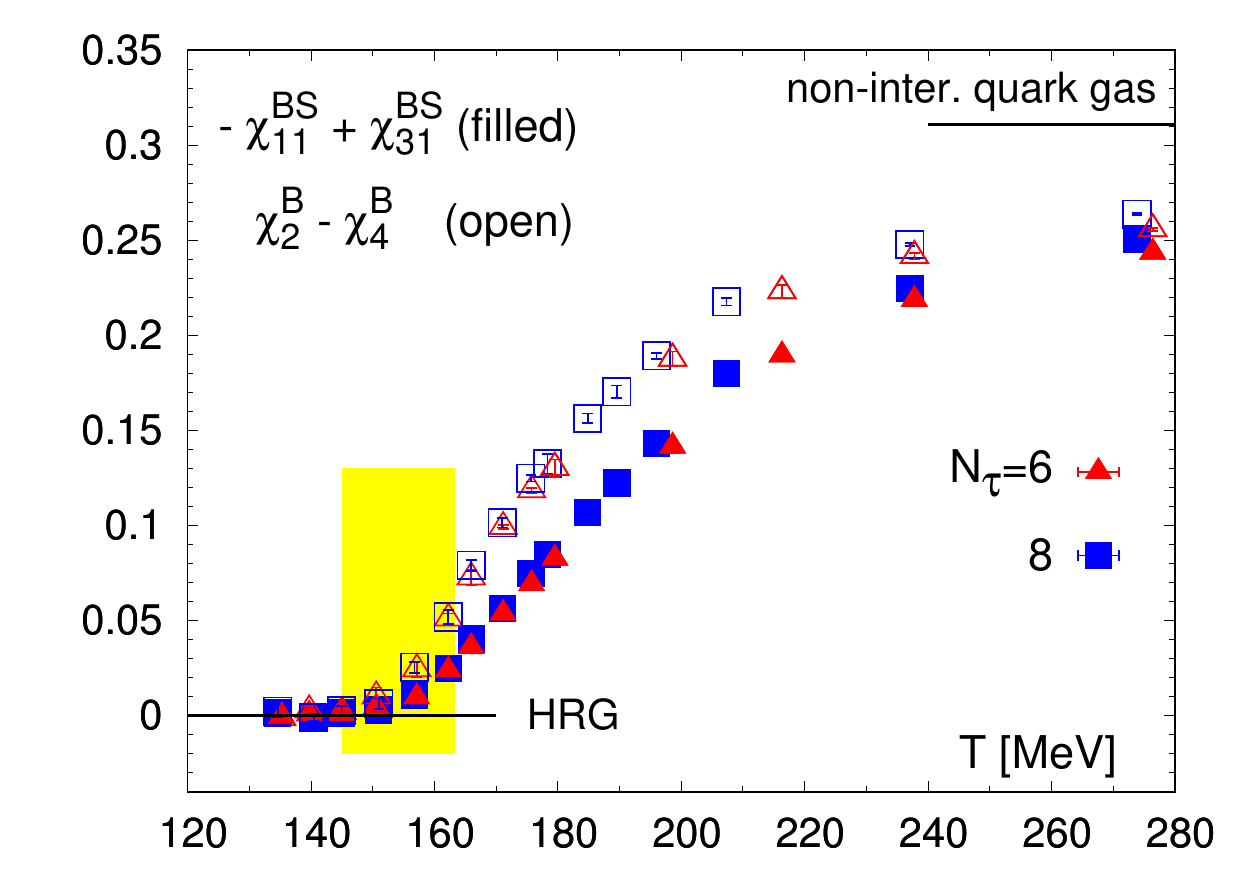}\hspace*{-0.7cm}
\includegraphics[width=68mm]{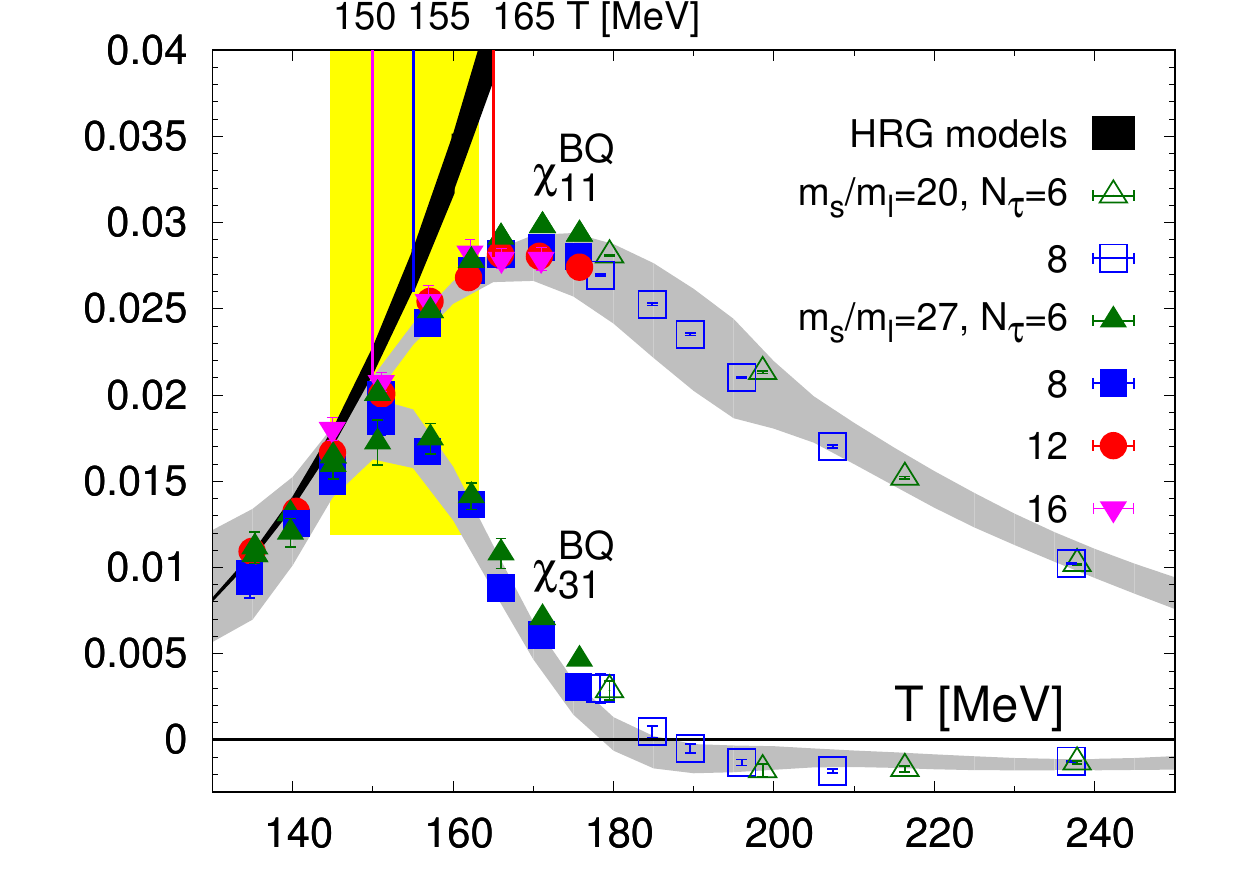}
\caption{{\it Left:}
The difference of second and fourth order cumulants of net baryon-number
fluctuations and their correlation with net strangeness fluctuations.
\newline
{\it Right:}
Second and fourth order cumulants of correlations between moments of
net baryon-number and electric charge fluctuations.
\vspace*{-0.5cm}
}
\label{fig:chi42}
\end{center}
\end{figure}

Results for lines of constant physics and the crossover line in the 
$T$-$\mu_B$ phase diagram of QCD are shown in Fig.~\ref{fig:LCP}~(right).
Also shown in this figure are experimental results for sets of freeze-out
parameters, $(T_f,\mu_B^f)$, determined by the ALICE Collaboration 
\cite{Floris:2014pta} at the
LHC and the STAR Collaboration \cite{Adamczyk:2017iwn}
at different beam energies at RHIC. These
parameter sets have been obtained by comparing experimentally determined
hadron yields with those calculated in HRG models.
Unfortunately, at present results from both collaborations differ 
significantly at large collision energies. 
Also shown in Fig.~\ref{fig:LCP}~(right) are results for the hadronization 
temperature extracted by Becattini et al. \cite{Becattini:2016xct}. 
The latter analysis, unlike the STAR data, leads to hadronization
parameters that nicely follow LCPs
and have a curvature consistent with the QCD crossover line. However,
the resulting hadronization temperatures are large and
%
difficult to reconcile with the chiral 
transition temperature, $T_c=154(9)$~MeV. 
They seem to suggest that hadronization takes place 
at $T$-values,
where QCD thermodynamics already shows many features of 
a partonic medium, albeit not a free gas of quarks and gluon, and is no 
longer compatible with HRG thermodynamics. This is apparent from the
temperature dependence of $2^{nd}$ and $4^{th}$ order cumulants
shown in Fig.~\ref{fig:chi42}. 
Obviously the agreement between cumulants that
only differ by two derivatives with respect to $\mu_B$
breaks down for $T\gsim 155$~MeV. As suggested in
\cite{Bazavov:2013dta} this may hint at the appearance of quasi-particles
with baryon number $B\ne \pm 1$, or at least the importance of interactions,
which also may be interpreted as the appearance of clusters with
$B\ne \pm 1$.

The $2^{nd}$ and $4^{th}$ order cumulants shown in Fig.~\ref{fig:chi42}
also are the first terms in Taylor expansions of net conserved
charges\footnote{For simplicity we set $\mu_Q=\mu_S=0$.}, e.g.
\begin{eqnarray}
n_B &=& \chi_2^B \frac{\mu_B}{T} +\frac{1}{6}\chi_{4}^{B}
\left( \frac{\mu_B}{T}\right)^3 + {\cal O}(\mu_B^5) \; ,
\nonumber \\
n_X&=&\chi_{11}^{BX} \frac{\mu_B}{T} +\frac{1}{6}\chi_{31}^{BX}
\left( \frac{\mu_B}{T}\right)^3+
{\cal O}(\mu_B^5) \; ,\; X=S,\ Q \;\; .
\label{charge}
\end{eqnarray}
This suggests that even on the level of particle yields
systematic differences between QCD and HRG model calculations
will show up at temperatures larger than $T\simeq 155$~MeV and will
become more pronounced for $\mu_B\ne 0$.

\begin{figure}[t]
\begin{center}
\hspace*{-0.4cm}\includegraphics[width=68mm]{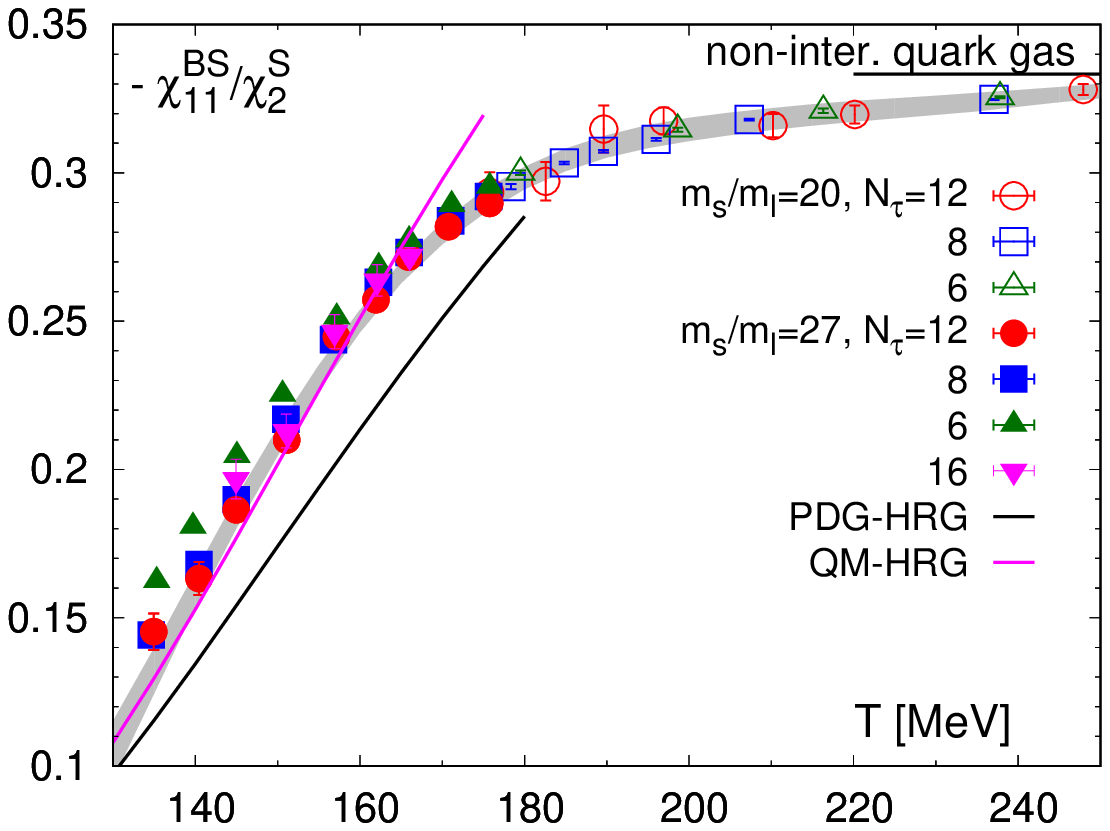}\hspace*{-0.7cm}
\includegraphics[width=68mm]{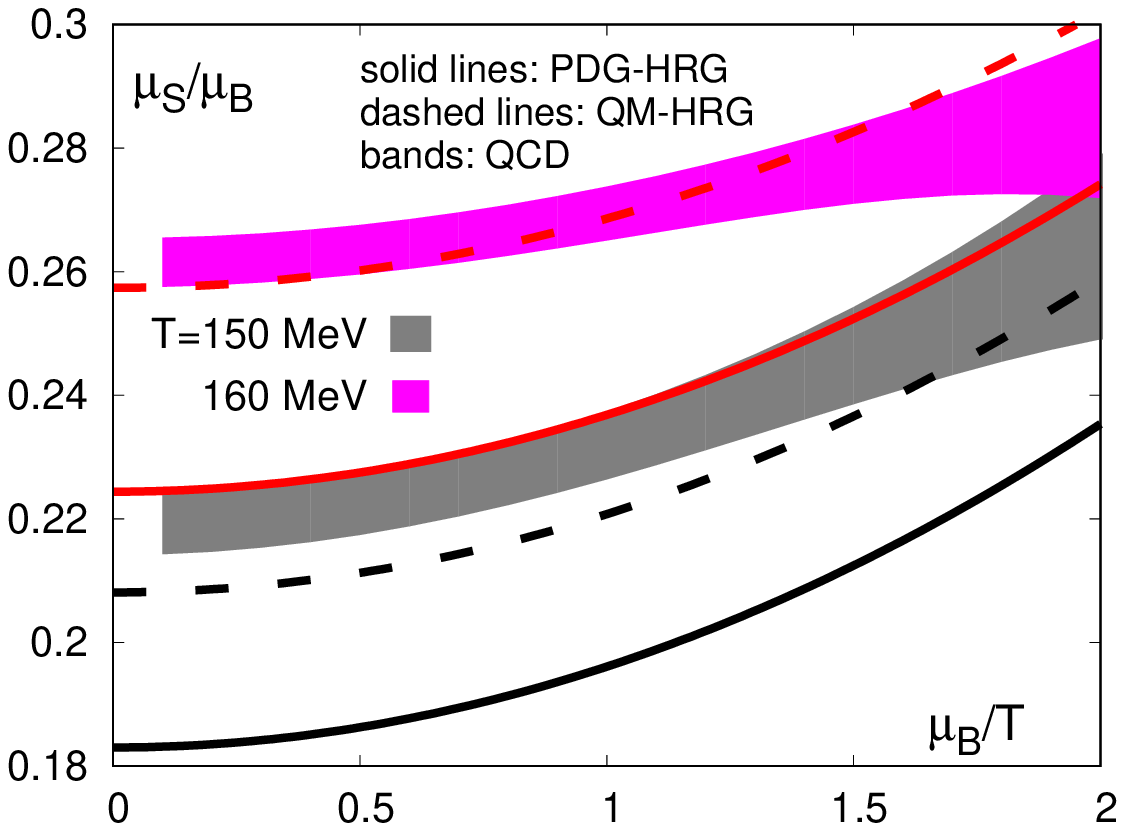}
\caption{{\it Left:}
The correlation between net baryon-number and strangeness fluctuations
normalized to the variance of strangeness fluctuations (see text for details).
\newline
{\it Right:} Ratio $\mu_S/\mu_B$ needed to adjust
$n_S=0$ and $n_Q/n_B=0.4$ in QCD. Lines
show corresponding results for two different versions of HRG
models (see text).
}
\label{fig:cumulants}
\end{center}
\end{figure}

Moreover, even at temperatures where hadrons are the dominant degrees of
freedom and HRG models may be appropriate to describe the thermodynamics
of strong interaction matter, such a description is sensitive to the
particle content in the HRG. In particular, it has been noted that
systematic differences show up in the strangeness sector
\cite{Bazavov:2012vg}. 
The correlation
between net baryon-number and strangeness fluctuations, $\chi_{11}^{BS}$, 
is systematically larger in QCD than in HRG model calculations that are based only on 
experimentally known hadron resonances (PDG-HRG). In fact, at 
temperatures $T\lsim 160$~MeV $\chi_{11}^{BS}$ is in good agreement with
HRG model calculations that also include resonances predicted to exist
in quark models (QM-HRG). This is shown in Fig.~\ref{fig:cumulants}~(left).
A consequence of this difference is that values of $\mu_B$ and/or $T$
that describe identical thermal conditions, e.g. 
identical values of the charge fluctuations, do not match one-to-one
between HRG 
and QCD calculations. Fig.~\ref{fig:cumulants}~(right) shows 
the values for the ratio $\mu_S/\mu_B$
needed to ensure strangeness neutrality, $n_s=0$, in strong interaction
matter. It is evident that a certain ratio $\mu_S/\mu_B$ corresponds to
temperatures that differ by about $10$~MeV in QCD and HRG models.
Similar information can be deduced from Fig.~\ref{fig:cumulants}~(left)
when considering a fixed ratio $\chi_{11}^{BS}/\chi_2^S$.

\section{Conclusions}

Differences between the modeling of strong interaction matter in terms
of HRG model thermodynamics and a QCD calculations rapidly become large
for $T\gsim 155$~MeV.
This easily can lead to a $\sim 10\%$ mismatch between 
temperature and chemical potential
values determined as freeze-out parameters in QCD and model calculations. 
The width of the 
singular region around a critical point may be of similar size. A $10$~MeV
accuracy on e.g. the freeze-out temperature $T^f(\mu_B)$ thus 
%
may decide whether or not critical behavior is detectable at
all through measurements of cumulants of conserved charge fluctuations.

\vspace*{0.2cm}
\noindent
{\it Acknowledgements:}
This work was supported in part through Contract No. DE-SC001270 with the
U.S. Department of Energy, the grant 05P12PBCTA of the German Bundesministerium f\"ur Bildung und Forschung and the grant 56268409 of the German Academic 
Exchange Service (DAAD).


\end{document}